\documentstyle[PASJadd,psfig]{PASJ95}
\draft

\begin{document}
\setcounter{page}{1}

\title{[OII]$\lambda 3727$ Emission from the Companion
to the Quasar BR 1202-0725 at $z$=4.7}

\author{Kouji {\sc Ohta}, Tsuyoshi {\sc Matsumoto}, and 
Toshinori {\sc Maihara} \\
{\it Department of Astronomy, Kyoto University, Sakyo-ku, Kyoto 606-8502} \\
{\it E-mail(KO): ohta@kusastro.kyoto-u.ac.jp}  
\\[6pt]
Fumihide {\sc Iwamuro}, Hiroshi {\sc Terada}, Miwa {\sc Goto}, Kentaro {\sc Motohara}, \\
Tomoyuki {\sc Taguchi}, and Ryuji {\sc Hata} \\
{\it Department of Physics, Kyoto University, Sakyo-ku, Kyoto 606-8502} \\
Michitoshi {\sc Yoshida}  \\
{\it Okayama Astrophysical Observatory, National Astronomical Observatory,} \\
{\it Kamogata-cho, Okayama 719-0232} \\
Masanori {\sc Iye} \\
{\it National Astronomical Observatory, Mitaka, Tokyo 181-8588} \\
and \\
Chris {\sc Simpson}, Tadafumi {\sc Takata} \\
{\it Subaru Telescope, National Astronomical Observatory of Japan,} \\
{\it Hilo, Hawaii 96720, USA}}

\abst{Results of a narrow-band imaging for the redshifted [OII]$\lambda
3727$ emission around a quasar at $z=4.7$ obtained with the Subaru
telescope and CISCO (a Cassegrain near infrared camera) are
presented.
A significant emission line is detected in the narrow-band H$_2$
($v=1-0$ S(1)) filter at a location $\sim$
2.$^{\prime \prime}$4\ northwest from the quasar, where the presence of a
companion has been reported in Lyman $\alpha$ emission and the rest-frame
UV continuum.
We identify this line as [OII]$\lambda 3727$ emission and
confirm that the source really is a companion at $z=4.7$.
The [OII]$\lambda 3727$ flux from the companion is estimated to
be $2.5 \times 10^{-17}$ erg s$^{-1}$ cm$^{-2}$.
If the companion is a star forming object, the inferred star
formation rate is as high as 45-230 $M_{\odot}$ yr$^{-1}$ even
without assuming the extinction correction.
This value is higher than those derived from the Lyman $\alpha$
emission or from the UV continuum.
Thus, provided that the difference is caused by dust extinction,
the extinction corrected star formation rate is
calculated to be 45 to 2300 $M_{\odot}$ yr$^{-1}$ depending on
the assuming extinction curves.}

\kword{Galaxies: formation --- Quasars: formation --- Quasars: individual 
(BR 1202$-$0725)}

\maketitle
\thispagestyle{headings}

\section
{Introduction}
There are growing numbers of observational and theoretical
suggestions that high redshift quasars are under intense star burst
and are related to an early phase of galaxy formation (e.g.,
Djorgovski et al. 1999; Haehnelt, Rees 1993).
BR 1202$-$0725 at $z$=4.7 is one of the most distant quasars
known to date and
provides us a unique opportunity for studying galaxy formation
process and its relationship to the quasar activity.
The quasar has an optical (rest UV continuum) companion located at
$\sim $2.$^{\prime \prime}$4  northwest from the quasar and 
it is also seen in  Lyman $\alpha$ emission (Djorgovski 1995; Hu et al. 1996; Petitjean et
al. 1996; Fontana et al. 1996).
The projected distance between the companion and the quasar is
$\sim 14$ kpc.
(We use a cosmological parameter set of q$_0 =0.5$ and $H_0 = 50$
km s$^{-1}$ Mpc$^{-1}$ throughout this paper.)
The flux of the Lyman $\alpha$ emission is estimated to be
$(1.5 \sim 2.5) \times 10^{-16}$ erg s$^{-1}$ cm$^{-2}$
(Hu et al. 1996; Petitjean et al. 1996).
If the Lyman $\alpha$ emission is attributed to star formation,
an expected star formation rate is 28-47 $M_{\odot}$ yr$^{-1}$.
A high angular resolution image in $I_{814}$-band taken with the
Hubble Space Telescope (HST) shows
that the companion has an elongated structure toward the quasar;
it may be tidally interacting with the quasar and may be merging into
the host galaxy (Hu et al. 1996).

Another important aspect in this system is the presence of a
huge amount of molecular gas and dust.
CO emission lines from this system is detected by Ohta et al. (1996)
and Omont et al.  (1996b) and the dust thermal emission by
McMahon et al. (1994), Isaak et al. (1994), Omont et al. (1996a),
and more recently by Benford et al.  (1999).
The estimated mass of the molecular gas amounts up to $\sim
10^{11} M_{\odot}$,  which is comparable to a stellar mass of a
present-day giant elliptical galaxy.
Presence of such an amount of molecular gas implies that
the system is considered to be under intense star burst 
($\sim 10^3 M_{\odot}$ yr$^{-1}$), i.e., 
under a forming phase in the early universe.
Distributions of the dust and the molecular gas show a double peak
structure;
1.35 mm continuum and CO emission show a peak at the quasar position and
at $\sim 4$$^{\prime \prime}$\  northwest of the quasar (Omont et al. 1996b;
Kawabe et al. 2000).
The intensity of the emission is roughly comparable with each
other.
There is no optical counterpart at the northwestern
position; the optical companion is located between the quasar
and the dust/molecular companion.
A recent high angular resolution radio continuum map obtained
with the Very Large Array also shows the double peak
structure at 1.4 GHz (Yun et al. 2000), and the star formation rate
is estimated to be again $\sim 10^3  M_{\odot}$ yr$^{-1}$, by
assuming that the radio continuum emission traces star formation activity
which is inferred from the fact that the spectral energy distribution
longward of far-infrared is quite similar  to that of the
central region of M 82 (Kawabe et al. 1999; Yun et al. 2000).

The Lyman $\alpha$ emission from the companion of the quasar is probably
quenched very much not only due to dust extinction at UV wavelength
but also due to the resonance scattering, thus emission lines at longer
wavelength are more suitable to trace star formation activity.
Further, a number of high redshift Lyman $\alpha$ emitters have been
found (e.g., Pascarelle et al. 1996; Hu \& McMahon 1996;
Cowie \& Hu 1998, Thommes et al. 1998),
but the degree of extinction in these systems is hardly examined.
Thus the determination of fluxes of other emission lines in 
this system would provide us an insight for the amount of dust
obscuration and a more reliable star formation rate for these
Lyman $\alpha$ emitters.

The [OII]$\lambda 3727$ emission from the companion of BR1202-0725
is redshifted to 2.123$\mu$m, which lies very close to the rest
wavelength of H$_2$ ($v=1-0$ S(1)) emission.
We used hence a narrow-band filter for this line to produce
a monochromatic [OII]$\lambda 3727$ image of the BR1202-0725,
using  the Subaru telescope and CISCO
(near infrared imaging spectrograph at the Cassegrain focus).
Here we report the detection of [OII]$\lambda 3727$ emission
line from the companion and possible other companions. 

\section
{Observations and Data Reduction}
Observations were made with CISCO (Motohara et al. 1998) attached to the
Cassegrain focus of the Subaru telescope  on 1999 March 4,
April 1, and April 2, during a period of
test observing run for the instrument and the telescope.
Since the weather condition was not good on March 4, we used
only the data taken in April.
Detector used was a 1k by 1k HgCdTe array with the pixel scale of
0.$^{\prime \prime}$116, giving a field of view of $\sim 2^{\prime}$.
Narrow-band [OII] line imaging with a filter centered at 
2.1196$\mu$m with a FWHM of 0.0199 $\mu$m (the transmission between
2.113$\mu$m and 2.126$\mu$m ranges from 80\% to 88\%) at 77 K and
$K^{\prime}$ ($1.96 - 2.30 \mu$m) imaging were carried out. 
The images were taken with an octagonal dither pattern of which 
diameter is $\sim 20^{\prime \prime}$,  and 
six frames were taken at each telescope pointing.
The offsets were introduced to reduce non-uniformity of the
pixel sensitivity.
The exposure time for each frame  was 50 sec for the narrow-band
imaging and 10 sec or 20 sec for the $K^{\prime}$ imaging.
The total exposure times were 1200 sec on Apr. 1 and 2400 sec on
Apr. 2 for the narrow-band imaging, and 480 sec both on Apr. 1 and Apr.
2 for the $K^{\prime}$ imaging.
Seeings during the observations were $\sim 0.^{\prime \prime}$6 to 
$\sim 1.^{\prime \prime}$0 (FWHM) and
$\sim 0.^{\prime \prime}$3 to $\sim  0.^{\prime \prime}$6 (FWHM)
on Apr. 1 and Apr. 2, respectively.

Data reduction was made with IRAF\footnote{IRAF is distributed
by the National Optical Astronomy Observatories, which is
operated by the Association of Universities for Research in
Astronomy, Inc. under cooperative agreement with the National
Science Foundation.}.
For each frame, the frame taken at the next pointing of the
telescope was subtracted;
since six images were taken at one pointing position,
for the n-th frame in a pointing, the n-th frame in the next
pointing was subtracted.
For the last pointing data, we subtracted the frames taken before that
pointing.
The small scale dark pattern was eliminated by this procedure.
The residual of the background shows a smooth distribution
in each quadrant of the detector.
Thus we fitted a two dimensional polynomial function of fourth order
in $x$-direction and third in $y$-direction with cross terms to the
residual background in each quadrant, and subtracted it.
The background of the images became flat and zero at this stage,
except for the first 6 frames in each night presumably due to some
unstable condition of the readout of the detector at that time.
The standard $K^{\prime}$ sky frame made from many observing
runs was applied for flat fielding at this stage both for the
$K^{\prime}$ data and the narrow-band data, because the narrow-band
was actually inside in a part of the $K^{\prime}$-band.
To check the robustness of the result, we also examined the case
without applying the flat fielding for the narrow-band data.
The discrepancy between the two procedures is not significant,
but a flux of the companion discussed below is 8 \% smaller for the 
latter case.
It should be noted that the flat fielding error does not affect the
present results and discussion described below so much,
because we used the quasar as a calibrator and the global variation
in the flat frame must be insignificant 
between the quasar position and the companion position.
In addition, the pixel to pixel variation is reduced much for
the photometry since a relatively large (8 pixel diameter) aperture
was used in this study.

The images were then shifted and smoothed with a Gaussian kernel
to match the seeing size.
We discarded poorer frames with a seeing size (FWHM) larger than
8.57 pixel (0.$^{\prime \prime}$99) for the data taken on Apr. 1 and 4.98 pixel
(0.$^{\prime \prime}$58) for the data  taken on Apr. 2.
Finally the images were stacked to make the final image,
excluding the first 6 frames, and the resulting narrow-band image
obtained on Apr. 2 is shown in figure 1.
(In figure 1, a slight smoothing was applied for clearer
presentation.)
The total effective integration times for the narrow-band imaging
are 900 sec for Apr. 1 data and 1900 sec for Apr. 2 data.
The corresponding effective integration time for the broad band
filter was 420 sec both for Apr.  1 and Apr. 2 data.

In figure 1, the companion is seen at $\sim 2.^{\prime \prime}4$ 
northwest of the quasar.
In order to examine the reality of the presence of the companion
in the narrow-band image, we divided the data (Apr. 2 data) into
the first half and the last half.
Although the resulting combined images 
have lower signal-to-noise ratios, the feature is still seen in
both images.
For a further test, combined images were constructed by discarding
the frames which were obtained at a telescope pointing in
order to avoid influence by some sprious feature at a particular
position in the detector (e.g., hot pixel).
For the Apr. 2 data, since we used seven pointing data for figure 1,
we combined six  pointing data into one image.
The resulting seven images are displayed in figure 2 without smoothing.
In every image, the companion is detected significantly, thus a
sprious feature at a particular
position in the detector does not produce the companion feature.
The FWHM of the companion feature measured by Gaussian fitting
is 10 to 20 \% larger or smaller
than that of the quasar in each image of figure 2.
The difference is considered to be attributed to the low
signal-to-noise ratio.

Another possible cause for a fake companion is a ghost of the quasar.
If the companion feature was the ghost, its position relative to
the quasar depends on the position of the object in the detector,
i.e., the telescope pointing.
We combined the frames taken in the western and eastern regions of
the detector with respect to its center, respectively.
Twenty one frames taken in the eastern region in the detector were
combined in one image and is shown in the left panel of figure 3.
Similarly, seventeen frmaes taken in the western region was
combined into one image and is shown in the right panel of figure 3.
In both images, the companion feature is located at 
2.$^{\prime \prime}4 \pm 0.^{\prime \prime}1$ northwest of the quasar.
Thus we conclude the companion feature is not the ghost image.

We also reduced the data independently using another manner of
the data reduction which is adopted by Iwamuro et al. (2000) for
near-infrared observations of the Hubble Deep Field North with
the CISCO.
Although the morphology of the feature is slightly different, the
existence of the companion in the image is as clear as that in
figure 1.
A count rate for the companion is differet from that obtained above
by about 10\%.
Therefore we conclude that the companion feature is real.

The flux calibration for the narrow-band imaging was made using the
filter transmission curve under the cooled condition and
a $K$-band spectrum of the quasar BR 1202-0725 taken with the
CGS-4 in the UKIRT service observing run in 1996.
The slit width used was 1.$^{\prime \prime}$23 and total exposure time was 32 min.
Flux calibration was made using UKIRT standard BS4533 observed in
the same observing run.
The $K$-band spectrum shows only a continuum and
the mean continuum flux level agrees well within $\sim 15$ \% with the
$K^{\prime}$ magnitude obtained by Hu et al. (1996) and no 
significant emission lines is seen in the filter transmission.
The flux calibration was done by comparing count rates of the
quasar and the companion; the uncertainty of the count rate
ratio estimated though figure 2 is about 10 \%.
We suppose that the origin of the feature in the narrow-band image is 
[OII]$\lambda 3727$  emission as described below, and assume that
the shape and the redshift of the [OII] emitter  are the same as
those for the Lyman $\alpha$ emission obtained by Petitjean et al.  (1996).

\section
{Results}
\subsection
{The northwestern companion}
As seen in the narrow-band image of figure 1,
the highest peak position of the companion at the northwest of
the quasar is located at 2.$^{\prime \prime}$4 from the quasar,
which is different from the reported position of the Lyman
$\alpha$ emission region by 0.$^{\prime \prime}$2, but the entire position
of the companion is considered to agree well with those
of the Lyman $\alpha$ companion and the optical continuum feature.
(The position of the Lyman $\alpha$ companion coincides with
that of the continuum feature (Hu et al. 1997).)
The extent of the companion is 1$^{\prime \prime}$\  to 
2$^{\prime \prime}$ and the feature is concentrated within
a 0.$^{\prime \prime}$5   region.
The elongated structure seen in the HST $I$-band image is not seen,
though the signal-to-noise ratio is not sufficient to discuss
the shape of the companion.
The total counts (ADU) of the companion in figure 1 in an aperture
with a diameter of 8 pixels (0.$^{\prime \prime}$93) is $\sim 110$ with
a peak value of $\sim 4$ counts.
A fluctuation of total counts in the same aperture evaluated in 
blank fields around the quasar is $\sim 20$ counts with a mean of
about 0.
Thus the significance of the feature is estimated to be about 5 sigma. 
The data taken on Apr 1, however, do not show the presence of
this feature significantly, because of the shorter exposure time
and the poorer seeing condition.
In fact, the image constructed from the data taken on Apr. 2
by adjusting the exposure time to 900 sec and seeing size to
0.$^{\prime \prime}$99 does not show the  significant feature of the companion.
(Thus we did not combine the data taken on Apr. 1.)
Our $K^{\prime}$-band image is not deep enough to detect the
companion; the $K$-band flux density of the companion is $1.9 \times
10^{-20}$ erg s$^{-1}$ cm$^{-2}$ \AA$^{-1}$ (Hu et al. 1996) and 
is lower than our detection limit.

The flux of the companion feature in the narrow-band image
(in the circular aperture of 8 pixels or 0.$^{\prime \prime}$93)
is $2.5  \times 10^{-17}$ erg s$^{-1}$ cm$^{-2}$,
while the contribution from the continuum is about 10 times smaller than
this value as mentioned above.
Therefore, considering the good positional coincidence to the
Lyman $\alpha$ emitter and the continuum features, we identify
the companion seen in the narrow-band image with the [OII]$\lambda
3727$ emitter at $z=4.7$.
The estimated observed equivalent width for the [OII]$\lambda 3727$
emission is $\sim 1300$ \AA, that corresponds to the rest equivalent
width of $\sim 230$ \AA.
The obtained [OII]$\lambda 3727$ flux is slightly lower than the upper limit set by Pahre and
Djorgovski (1995) obtained with the NIRC (near infrared camera)
at the Keck telescope.
The non-detection with the NIRC may be due to a combination of
the lower transmission of the narrow-band filter
for H$_2(1-0$) line (Teplitz et al. 1998), 
a shorter integration time (1080 s), and a larger seeing size
(typically 0.$^{\prime \prime}75$).

\subsection
{The southwestern companion and other companion
candidates}
In the $I$-band image obtained  with the HST, there is another extended
optical feature pointing radially to the quasar at $\sim
3$$^{\prime \prime}$\  southwest from the quasar (Hu et al. 1996).
No significant emission was detected toward this SW companion.
The $3\sigma$ upper limit on this position is $\sim 1.4 \times 10^{-17}$ erg
s$^{-1}$ cm$^{-2}$ with the aperture of 8 pixels (0.$^{\prime \prime}$93),
provided that the velocity feature is the same as that of the Lyman
$\alpha$ emission of the northwestern companion.
Hu et al. (1997) reported a detection of Lyman $\alpha$ emission
toward this companion as well as [OII]$\lambda 3727$ emission which was
observed with UH 1k by 1k IR camera attached to the Canada-France-Hawaii
Telescope. (They did not present the flux data.)
Our result does not confirm the [OII] emission toward the SW
companion.
Since no significant dust thermal emission, CO emission, and
radio continuum emission are detected at this position, 
the object might not be placed at $z=4.7$ or be a chemically unevolved
and less extincted object with a relatively low intrinsic star
formation rate.

In figure 1, there may be other [OII]$\lambda 3727$ emission regions
around the quasar:
at $\sim$ 1.$^{\prime \prime}$5 south-by-east and 
$\sim$ 1.$^{\prime \prime}$5 south-by-west from the quasar.
Although these features are also seen in figures 2 and 3, they are very
close to the quasar and might be outskirts of the point spread function.
There might be another companion at $\sim$ 1$^{\prime \prime}$\ to 
2$^{\prime \prime}$\ south from the companion.
If these features are real, their close projected distances from the
quasar of about 10 to 20 kpc suggest that they may be subgalactic
clumps under active star formation merging into a galaxy,
the host of the quasar.
Alternatively, they might be gas clumps ionized by the quasar that
are merging into the host galaxy.
However, since all these features have the low signal-to-ratios,
deeper imaging and spectroscopic observations are obviously
needed to examine the existence of them  and to discuss these
possibilities.

\section{Discussion}
There are alternative interpretations for the origin of the emission
line in the northwestern companion:
one is a star forming object and the other one is a photoionized gas
illuminated by the quasar.
Petitjean et al. (1996) discussed both possibilities 
by constructing model continuum spectra with emission lines.
In the star forming object hypothesis, they assumed a star
formation rate of 13 $M_{\odot}$ yr$^{-1}$ and 0.1 solar abundance 
without dust extinction.
(They used q$_0 =0.5$ and $H_0 = 75$ km s$^{-1}$ Mpc$^{-1}$.)
A shortcoming for this model  arises from
the fact that contribution from Lyman $\alpha$ emission in the model
is smaller than the observed value to account for intensities of
continuum and emission lines simultaneously.
In the photoionization model, the model $V$-band flux is too small
and the Lyman alpha emission is too strong (about one order of
magnitude larger than the observed value).
In the latter model, a strong NV$\lambda 1240$ and CIV$\lambda 1549$ 
emission line should be seen in the spectrum of the companion.
However the spectrum of the companion by Petitjean et al. (1996) seems
not to show the NV$\lambda 1240$ emission, and Hu et al. (1997)
reported the absence of CIV$\lambda 1549$ emission line in the companion.
Thus the identification of the northwestern companion with a star
forming object seems to be more plausible.

The star formation rate estimated from the [OII]$\lambda 3727$ emission
luminosity is 45 to 230 $M_{\odot}$ yr$^{-1}$ depending on the adopted
conversion factor from the [OII]$\lambda 3727$ luminosity to
the H$\alpha$ luminosity (Gallagher et al. 1989; Kennicutt 1992) 
without extinction correction.
The star formation rate estimated from Lyman $\alpha$ luminosity
is 28-47  $M_{\odot}$ yr$^{-1}$  under the case B recombination assumption
($L$(Lyman $\alpha$)/$L$(H$\alpha) = 8.7$) (Hu et al. 1996;
Petitjean et al. 1996).
The star formation rate estimated from the rest UV continuum (around
1400 \AA) (Madau et al. 1998) also gives a value of 22 $M_{\odot}$ yr$^{-1}$.
The star formation rate derived from the [OII] emission is 
larger than those estimated from the Lyman $\alpha$ emission and the
UV continuum by a factor of up to $\sim 10$.

The origin of the discrepancy must be due to the severer dust
extinction to the Lyman $\alpha$ emission and the UV continuum.
By assuming that the extinction corrected star formation rate
derived from the [OII] luminosity is equal to that from the Lyman 
$\alpha$ luminosity, the amount of extinction in the companion can be
estimated.
The estimated visual extinction ($A_V$) ranges from $\sim 0$ to 1.3
mag for the Milky Way type extinction curve (Scheffler and
Els\"asser 1988), $\sim 0$ to 0.5 mag for the SMC type extinction curve
(Calzetti et al. 1994), and $\sim 0$ to 1.1 mag for the Calzetti type
extinction curve (Calzetti 1997), by assuming $A_V = 3 E(B-V)$.
Thus the extinction at [OII]$\lambda3727$ emission is estimated
to be $\sim 0$ to 2.5 mag, and the
extinction corrected star formation rate amounts to 45 to 2300
$M_{\odot}$ yr$^{-1}$.
The emission line region is adjacent to the thermal dust emission
and  the CO emission line region which must also be an intense star
forming region at which the star formation rate inferred from the
CO luminosity and the radio continuum luminosity is at
the order of 1000 $M_{\odot}$ yr$^{-1}$.
The [OII] emitting region may be a less obscured region of the
starburst object.
However, the huge value of the estimated star formation ratio
($\sim 2000 M_{\odot}$ yr$^{-1}$) for the companion alone  may
indicate that at least a part of the [OII] emission does not come from
star formation, but from the ionized gas illuminated by the quasar.
Deep spectroscopic observations of the companion are required to
examine a fraction of the contribution from the ionization by
the quasar.

\par
\vspace{1pc}\par
It is our great pleasure to thank Subaru telescope team for
their enormous efforts for the construction of the telescope.
K.O. is supported by grant-in-aid from the Ministry of
Education, Science, Sports and Culture of Japan (117401230).

\clearpage

\section*{References}
\small

\re
Benford D.J., Cox P., Omont A., Phillips T.G., McMahon R.G.\
1999, ApJ 518, L65

\re
Calzetti D.\ 1997, AJ 113, 162 

\re
Calzetti D., Kinney A.L., Storchi-Bregmann T.\ 1994, ApJ 429, 582

\re
Cowie L.L., Hu E.M.\ 1998, AJ 115, 1319 

\re
Djorgovski S.G.\ 1995, in Science with the VLT, eds J.R. Walsh,
I.J. Danziger, (Springer Verlag, Berlin) p351

\re
Djorgovski S.G., Odewahn S.C., Gal R.R., Brunner R., Carvalho
R.R.\ 1999, in Photometric Redshifts and the Detection of High
Redshift Galaxies, eds R. Weymann, L. Storrie-Lombardi, M. Sawicki,
R. Brunner, (Astronomical Society of Pacific, San Francisco) in
press 

\re
Fontana A., Cristiani S., D'Odorico S., Giallongo E., Savaglio S.\
1996, MN 279, L27 

\re
Gallagher J.S., Bushouse H., Hunter D.A.\ 1989, AJ 97, 700

\re
Haehnelt M., Rees M.J.\ 1993, MN 263, 168

\re
Hu E.M., McMahon R.G.\ 1996, Nature 382, 231

\re
Hu E.M., McMahon R.G., Egami E.\ 1996, ApJ 459, L53 

\re
Hu E.M., McMahon R.G., Egami E.\ 1997, in the Hubble Space
Telescope and the High Redshift Universe, eds N.R. Tanvir,
A. Arag\'{o}n-Salamanca, J.V. Wall, (World Scientific, Singapore), p91

\re
Isaak K.G., McMahon R.G., Hills R.E., Withington S.\ 1994, MN 269, L28 

\re
Iwamuro F., Motohara K., Maihara T., Iwai J., Tanabe H., Taguchi
T., Hata R., Terada H., et al.
2000, PASJ in press

\re
Kawabe R., Kohno K., Ohta K., Carilli C.\ 1999, in Highly
Redshifted Radio Lines, eds. C.L. Carilli, S.J.E. Radford, K.M.
Menten, G.I. Langston, (Astronomical Society of Pacific,
San Francisco), p48

\re
Kawabe R., Kohno K., Ohta K., Tutui, Y., Yamada T., Carilli C. \ 2000 in
preparation

\re
Kennicutt R.C. Jr\ 1992, ApJ 388, 310

\re
Madau P., Pozzetti L., Dickinson M.\ 1998, ApJ 498, 106

\re
McMahon R.G., Omont A., Bergeron J., Kreysa E., Haslam
C.G.T.\ 1994, MN 267, L9

\re
Motohara K., Maihara T., Iwamuro F., Oya S., Imanishi M., Terada
H., Goto M., Iwai J., et al.\ 1998, Proc. SPIE 3354, 659

\re
Ohta K., Yamada T., Nakanishi K., Kohno K., Akiyama M., Kawabe R.\ 1996,
Nature 382, 426  

\re
Omont A., McMahon R.G., Cox P., Kreysa E., Bergeron J.,
Pajot F., Storrie-Lombardi L.J.\ 1996a, A\&A 315, 1

\re
Omont A., Petitjean P., Guilloteau S., McMahon R.G., Solomon P.M., 
P\'econtal E.\ 1996b, Nature 382, 428 

\re
Pahre M.A., Djorgovski S.G.\ 1995, ApJ 449, L1

\re
Pascarelle S.M., Windhorst R.A., Keel W.C., Odewahn S.C.\
1996, Nature 383, 45

\re
Petitjean P., P\'econtal E., Valls-Gabaud D., Charlot S.\ 1996, 
Nature 380, 411 

\re
Scheffler R.W., Els\"asser H.\ 1988, Physics of the Galaxy and
Interstellar Matter (Berlin: Springer Verlag)

\re
Teplitz H.I., Malkan M., McLean I.S.\ 1998, ApJ 506, 519

\re
Thommes E.,  Meisenheimer R., Fockenbrock R., Hippelein H.,
R\"oser H.-J., Beckwith S.\ 1998, MN 293, L6

\re
Yun M.S., Carilli C.L., Kawabe R., Tutui Y., Kohno K., Ohta K.\
2000, ApJ 528, in press


\label{last}

\clearpage

\begin{figure}
\begin{center}
\leavevmode
\psfig{file=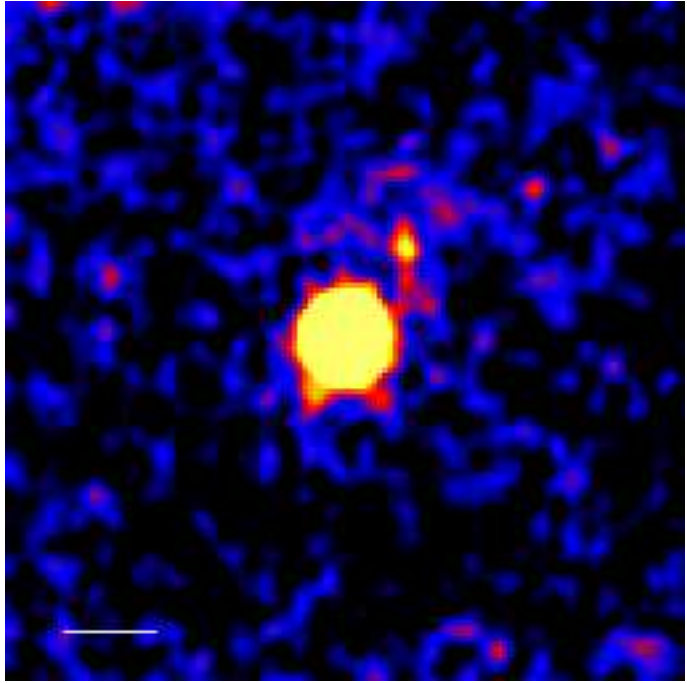}
\caption{
Narrow-band image for the redshifted [OII]$\lambda 3727$ emission
around the quasar BR1202-0725 at $z=4.7$.
Continuum emission is not subtracted.
North is at the top and east to the left.
A horizontal bar in the left bottom corner shows 2$^{\prime \prime}$ and the
displayed field of view is about 15$^{\prime \prime}$.
A central brightest object is the quasar.
The image is slightly smoothed with a Gaussian kernel for
clearer presentation; the FWHM of the quasar in the original image
is 0.$^{\prime \prime}$55 and that in the smoothed image is
0.$^{\prime \prime}$69.}
\end{center}
\end{figure}

\clearpage

\begin{figure}
\begin{center}
\leavevmode
\psfig{file=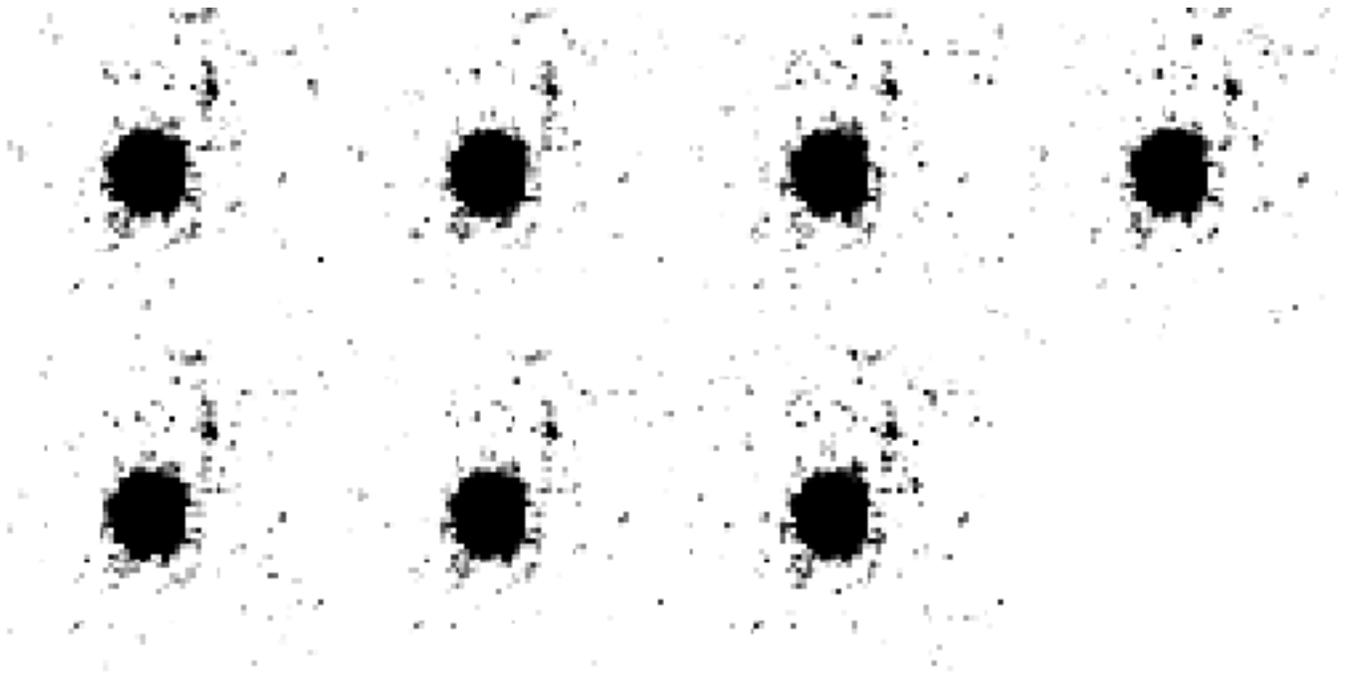}
\caption{
Narrow-band images constructed from the frames taken at six telescope
pointing.
Since we used seven pointing data for figure 1, seven images are
obtained.
North is at the top and east to the left.
No smoothing is applied to these images.
All the images show the presence of the companion feature
located at $\sim 2.^{\prime \prime}4$ northwest of the
quasar, which indicates the companion feature was not produced
by a spurious in a particular position of the detector.}
\end{center}

\clearpage
\end{figure}
\begin{figure}
\begin{center}
\leavevmode
\psfig{file=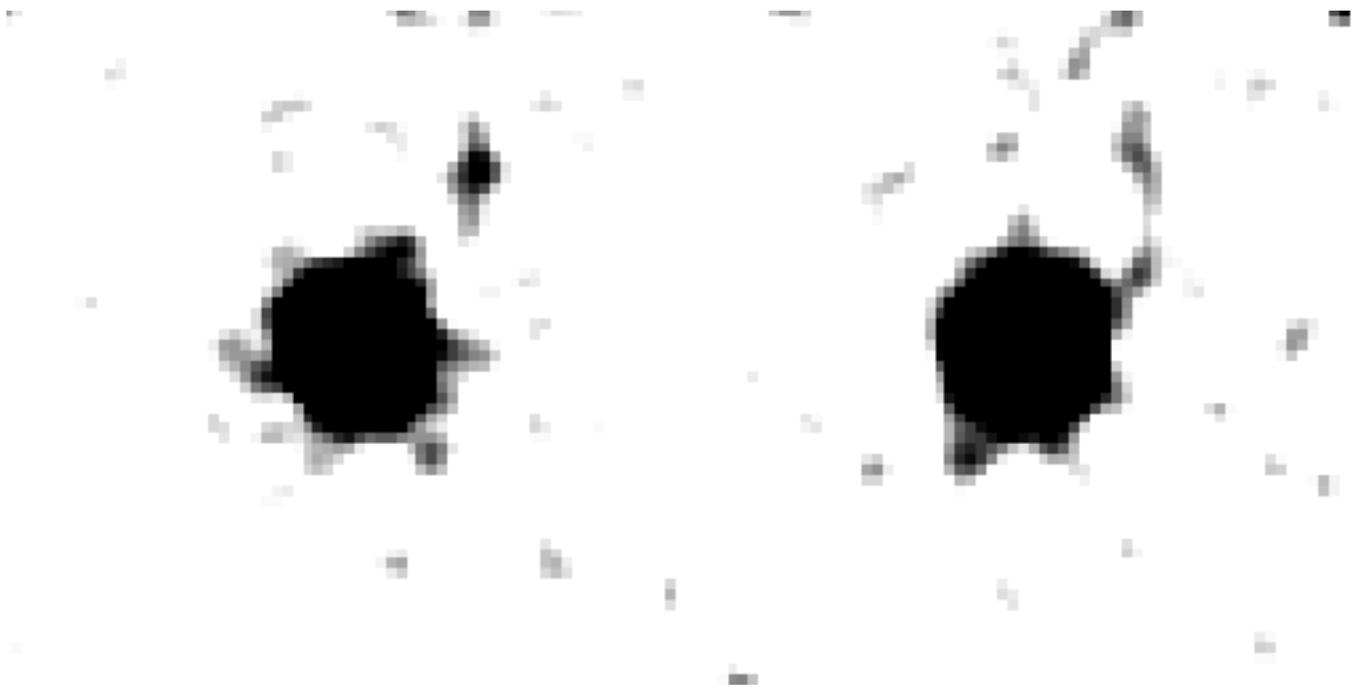}
\caption{
The left image is made from frames in which the object locates in
the eastern region of the detector, while the right image from
frames in which the object locates  in the western region of the
detector.
If the companion feature is a ghost of the quasar, the relative
position of the companion to the quasar should depend on the
object position in the detector.
In both images, the companion feature is located at $\sim
2.^{\prime \prime}4$ northwest of the quasar, and hence the
feature is considered not to be a ghost.
Total exposure times of the left and right images are 1050 s and
850 s, respectively.}
\end{center}
\end{figure}

\end{document}